\documentstyle[twoside,fleqn,espcrc2,epsf]{article}

\title{Recent results on light hadron and quark masses} 

\author{R.D. Kenway\address{Department of Physics and Astronomy, The
University of Edinburgh,\\ The King's Buildings, Edinburgh EH9 3JZ,
Scotland}}

\begin{document}

\begin{abstract}

Recent results for the spectrum of light hadrons provide clear evidence
for the failure of quenched QCD and encouraging signs that simulations
with dynamical sea quarks rectify some of the discrepancies, although
string breaking has not yet been observed. The use of perturbation
theory to match lattice quark masses to continuum schemes remains
questionable, but non-perturbative methods are poised to remove this
uncertainty. The inclusion of dynamical sea quarks substantially reduces
estimates of the light quark masses. New results for the lightest
glueball and the lightest exotic hybrid state provide useful input to
phenomenology, but still have limited or no treatment of mixing. The
$O(a)$-improved Wilson quark action is well-established in quenched QCD
for $\beta\geq 5.7$, with most parameters obtainable non-perturbatively,
in which range scaling violations are small. Progress has also been made
with high-order improvement schemes for both Wilson and staggered
quarks.  

\end{abstract}

\maketitle

\section{INTRODUCTION}

Calculation of the light hadron spectrum is an essential part of any
numerical simulation of QCD, because it provides the most direct way of
fixing the quark masses. The bare mass for each quark flavour is tuned
until the masses of correspondingly flavoured hadrons agree with
experiment, having chosen one dimensionful quantity to set the overall
scale. It is expensive to simulate at masses much below that of the
strange quark, so the $u$ and $d$ quark masses must be obtained by
chiral extrapolation. Electromagnetic corrections are ignored, so we
cannot distinguish the $u$ and $d$ masses.

The hadron spectrum is widely used to test scaling for improved lattice
actions; the improvement programme being our best hope for reliable
simulations with dynamical sea quarks. Once the known hadron masses have
been used to validate the simulations, the spectrum calculations provide
important predictions for phenomenology. They are guiding searches for
glueballs and for exotic mesons. Finally, the light quark masses are
some of the poorest-known Standard Model parameters and these
uncertainties matter; for instance, Standard Model predictions of
$\epsilon'/\epsilon$ are very sensitive to $m_s+m_d$~\cite{buras}. The
challenge is to find a reliable way of matching the lattice quark masses
to those in a continuum perturbative scheme, which can be used for
phenomenology.
 
\subsection{The determination of quark masses}

Quarks do not appear as asymptotic states in QCD, because of
confinement. As a consequence, it is usual to quote values for the
running quark mass at a particular scale in a specific scheme, eg
$m^{\overline{\rm MS}}(2\;{\rm GeV})$, although, alternatively, we could
run the mass up to high scales and quote the RG invariant (scheme
independent) value
\begin{equation}
M = \lim_{\mu\rightarrow\infty} m\,(2b_0g^2)^{-d_0/2b_0}
\label{eq:rg_invt_mass}
\end{equation}
where
\begin{eqnarray}
\mu\frac{\partial g}{\partial\mu} &=& \beta(g) = - b_0\,g^3+\ldots\\
\mu\frac{\partial m}{\partial\mu} &=& \tau(g)\,m = - d_0\,g^2m+\ldots
\end{eqnarray}
Given $M$, it is straightforward to run the mass down to any desired
scale, using continuum perturbation theory, provided we remain within
the perturbative regime.

For Wilson quarks, the pseudoscalar meson mass vanishes at
$\kappa=\kappa_{\rm crit}$ and the bare quark mass is 
\begin{equation}
m_qa = \frac{1}{2\kappa_q}-\frac{1}{2\kappa_{\rm crit}}
\end{equation}
where $\kappa_q$ is determined by tuning the mass of a $q$-flavoured
hadron to its experimental value. The renormalised quark mass may be
determined either using the vector Ward identity (conserved current),
\begin{equation}
\langle\partial_\mu V^a_\mu(x){\cal O}(0)\rangle
\hspace*{-0.3em}=\hspace*{-0.3em}
  \left[\frac{1}{2\kappa_2}-\frac{1}{2\kappa_1}\right]
  \hspace*{-0.4em}\frac{\langle S^a(x){\cal O}(0)\rangle(\mu)}{Z_S(\mu a)}
\end{equation}
so that
\begin{equation}
m_q^{\overline{\rm MS}}(\mu) = Z_S(\mu a)^{-1}\; m_q = Z_m(\mu a)\; m_q,
\end{equation}
or the axial Ward identity,
\begin{equation}
(m_1+m_2)(\mu) = \frac{Z_A}{Z_P(\mu a)}
                 \frac{\langle\partial_\mu A^a_\mu(x){\cal O}(0)\rangle}
                      {\langle P^a(x){\cal O}(0)\rangle(a)},
\label{eq:axialWI}
\end{equation}
which gives the renormalised quark masses without using $\kappa_{\rm
crit}$. Here, quantities denoted at the scale $\mu$ ($a$) are
renormalised (bare). For staggered quarks, the remnant chiral symmetry
ensures  $Z_A = 1$, $Z_P = Z_S = Z_m^{-1}$, so both methods are
essentially the same (unfortunately, the 1-loop term in the matching is
large, casting doubt on the reliability of perturbation theory).

\subsection{Some important questions}

In reviewing recent results for the light hadron spectrum, I will focus
on the following questions. Can we conclude that quenched QCD is wrong?
To what extent have we observed effects which can be attributed to
dynamical sea quarks? How effective are improved actions at reducing
scaling violations? Is the matching of lattice and continuum quark
masses under control? Finally, I will report on a few phenomenologically
important predictions for glueballs and hybrids.

\section{QUENCHED APPROXIMATION}

It has been known for some time that the quenched approximation gets the
spectrum in the strange sector wrong and underestimates hyperfine
splittings~\cite{yoshie97}. There are several related symptoms of this
failure, in particular, using $M_K$ and $M_\phi$ as input gives
different values for the strange quark mass.

\subsection{Chiral extrapolations}

Chiral extrapolation is now the main source of systematic error and a
key issue is whether the data requires the inclusion of quenched chiral
logarithms. If so, we need to be able to control the extrapolation to
within a few percent to show that the quenched approximation breaks down
in the $ud$ sector too. CP-PACS has reported such results for the
Wilson quark action at this conference, having increased the statistics
at its smallest lattice spacing during the
year~\cite{ruedi_talk,yoshie_talk}. As yet, though, this failure of the
quenched approximation has not been confirmed using staggered quarks, or
other improved actions.

In quenched QCD, the $\eta'$ meson remains light in the chiral limit,
because the infinite series of loop diagrams, which gives it a non-zero
mass in the full theory, is absent~\cite{quenched_chiPT}. The singlet
2-point function has a double pole, giving rise to so-called quenched
chiral logarithms, which diverge as $m_q\rightarrow 0$. The
renormalisation of quenched chiral perturbation theory has been carried
out at one loop~\cite{9708005}. Anomalous chiral behaviour persists in
dynamical simulations when the valence-quark and sea-quark masses are
different, as is necessarily the case for the strange quark in
simulations with $N_f=2$ sea quarks (partial quenching), and this can
produce strong dependence on the sea-quark mass when both it and the
valence-quark mass are small~\cite{part_quench}.

The CP-PACS data are consistent with the presence of quenched chiral
logarithms~\cite{ruedi_talk,yoshie_talk}. Both the pseudoscalar meson
masses and decay constants indicate a non-zero value for the coefficient
of the leading quenched chiral logarithm, $\delta=0.10(2)$. Also, using
quenched chiral perturbation theory to extrapolate the pseudoscalar
meson mass to zero, gives a value of $\kappa_{\rm crit}$ in better
agreement with a linear extrapolation of the axial Ward identity quark
mass, than a simple quadratic polynomial in the quark mass. 

\subsection{The quenched QCD spectrum}

\begin{figure}[htb]
\vspace*{25mm}
\epsfxsize=0.48\textwidth
\hspace*{-5mm}\epsfbox[10 60 510 560]{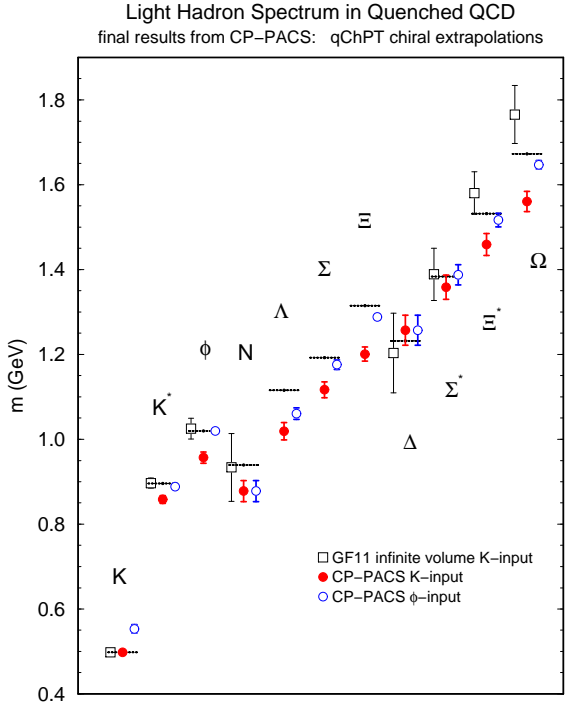}\\[-17mm]
\caption{CP-PACS results for the light hadron spectrum in quenched QCD
obtained using the Wilson quark action~\cite{ruedi_talk,yoshie_talk}.}
\label{fig:cppacs_quenched_final}
\end{figure}

The evidence of a non-zero value for $\delta$
justifies CP-PACS's use of quenched chiral perturbation theory to
produce their final results for the quenched spectrum. They obtain the
continuum limit, shown in Fig~\ref{fig:cppacs_quenched_final}, by linearly
extrapolating data at four lattice spacings, with $a^{-1}$ ranging from
2 to 4~GeV, on lattices which are all roughly 3~fm in size.

\begin{figure}[htb]
\vspace*{24mm}
\epsfxsize=0.46\textwidth
\hspace*{-3mm}\epsfbox[10 60 510 560]{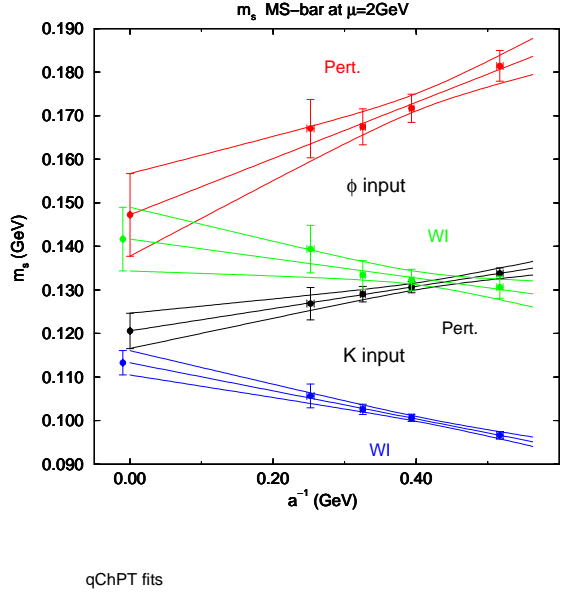}\\[-30mm]
\caption{CP-PACS results for the strange quark mass in quenched QCD,
defined using the vector (Pert in the figure) and axial (WI) Ward
identities~\cite{ruedi_talk,yoshie_talk}.}
\label{fig:cppacs_quenched_m_strange}
\end{figure}

Fig~\ref{fig:cppacs_quenched_final} confirms the picture that there is
no choice for the strange quark mass which can explain the whole
spectrum. Using $M_K$ as input, the meson hyperfine splitting, the octet
baryon masses, and the decuplet baryon mass splittings, are all too
small. Using $M_\phi$ to fix the strange quark mass reduces the
discrepancies amongst the baryon masses, but the other faults remain.
The problem is not due to the way in which the strange quark mass is
defined; as can be seen in Fig~\ref{fig:cppacs_quenched_m_strange}, the
vector and axial Ward identity definitions agree in the continuum limit,
but with limiting values which depend on the choice of input:
\begin{equation}
m_s^{\overline{\rm MS}}(2\mbox{ GeV}) = \left\{\begin{array}{cc}
              143(6)\mbox{ MeV} & (M_\phi\mbox{ input})\\
              115(2)\mbox{ MeV} & (M_K\mbox{ input})
             \end{array}\right.
\label{eq:cppacs_ms}
\end{equation}
The CP-PACS result for the $u$ and $d$ quark mass is
\begin{equation}
m_{ud}^{\overline{\rm MS}}(2\mbox{ GeV}) = 4.55(18)\mbox{ MeV}
\label{eq:cppacs_mud}
\end{equation}
again with consistency between the two definitions in the continuum
limit. The problem with the strange quark mass is directly related to
the low value of $J\equiv M_{K^\ast}({\rm d}m_{\rm V}/{\rm d}m_{\rm
PS}^2)$, since
\begin{equation}
\frac{(m_s)_K}{(m_s)_\phi}
\hspace{-1mm}=\hspace{-1mm} 
  \left[\left(\hspace*{-1mm}\frac{M_{K^\ast}}{M_K}\hspace*{-1mm}\right)^2
  \hspace{-2mm}\left(1+\frac{m_{ud}}{m_s}\right)\right]^{-1}\hspace*{-5mm}
  \frac{2J}{\frac{M_\phi}{M_{K^\ast}}-\frac{m_{\rm V}(0)}{M_{K^\ast}}}
\end{equation}
if linear chiral behaviour is assumed for $m_{\rm PS}^2$ and $m_{\rm
V}$~\cite{9801028}.

\begin{figure}[htb]
\epsfxsize=0.48\textwidth
\vspace*{-3mm}
\hspace*{-3mm}\epsffile{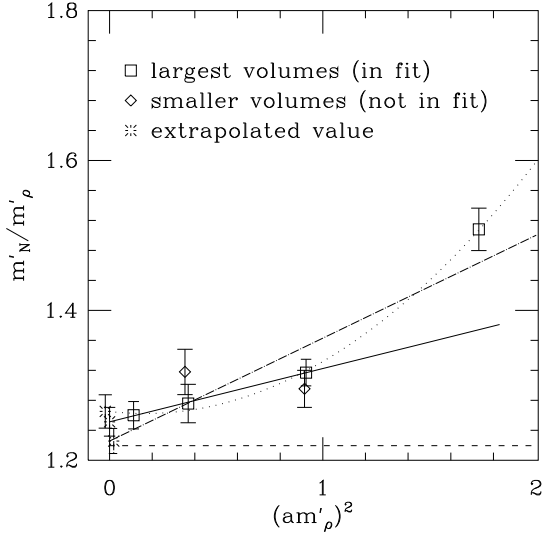}\\[-12mm]
\caption{Continuum extrapolation of MILC data for
$m_N/m_\rho$ in quenched QCD, using staggered 
quarks~\cite{9805004,gottlieb_poster}.}
\label{fig:milc_ratio}
\end{figure}
It is desirable to confirm the CP-PACS results using an improved action.
The most direct comparison of this sort is with MILC
results~\cite{9805004,gottlieb_poster} for staggered quarks, which are
also available on lattices of at least 3~fm in size. Chiral
extrapolation again turns out to be the most delicate issue. For
staggered quarks, it is the mass of the non-Goldstone pion which appears
in quenched chiral perturbation theory, and MILC includes terms linear
in this mass in the chiral extrapolation of the nucleon and vector
meson masses. The continuum extrapolation is shown in Fig~\ref{fig:milc_ratio}
and, combining various chiral and continuum fits, they quote the final
result
\begin{equation}
\frac{m_N}{m_\rho}=1.254\pm 0.018\pm 0.027,
\end{equation}
in agreement with the experimental value of 1.22. The statistical error is similar to that of CP-PACS, and the difference
between the results is only $2.5\sigma$, which is not a serious
disagreement. It would be interesting to combine the MILC data with that
of Kim and Ohta~\cite{kim_poster}, which is on a similar size lattice,
but with a smaller lattice spacing ($\beta=6.5$), to see if this error
can be reduced.

In summary, CP-PACS has provided us with evidence of the breakdown of
the quenched approximation for light hadrons at the few percent level,
the most striking symptom of this being non-uniqueness of the strange
quark mass, although confirmation from improved actions is still
awaited. 

\section{SEA-QUARK EFFECTS}

During the last few years, several groups have begun systematically to
explore the parameter space for simulations with dynamical quarks.
Finite-size effects must be checked, and this is not as straightforward
as in quenched QCD, because the lattice spacing at fixed sea-quark mass
depends on the physical quantity used to define it, and decreases
significantly with the sea-quark mass. So the volume is only
well-defined for chiral sea quarks.  

I will focus on simulations with two degenerate flavours of sea quark,
which may represent $u$ and $d$, in which case the strange quark must be
treated in the quenched approximation. The data for Wilson quarks,
obtained with unimproved and various choices of improved action
(SESAM~\cite{9806027}, CP-PACS~\cite{ruedi_talk,kanaya_talk} and
UKQCD~\cite{talevi_talk}), correspond to mass ratios $m_{\rm PS}/m_{\rm
V}$ between 0.6 and 0.8, and lattice sizes of around 2~fm, whereas the
MILC data for staggered quarks~\cite{gottlieb_poster} extend to a mass
ratio of around 0.3 on lattices which are somewhat larger. 

\subsection{Wilson quarks}

UKQCD has explored finite-size effects for hadron masses, in which
valence and sea-quark masses are the same, and finds that a significant effect
occurs at the lower end of the above range of quark masses, between
lattices sizes of 0.9 and 1.4~fm~\cite{talevi_talk}. Above 1.4~fm, no
effect is observed. Consequently, chiral extrapolations using linear and
quadratic polynomials in the valence and sea-quark masses, and including
cross-terms, is probably safe for the Wilson quark data mentioned above. 

SESAM notes that data at different sea-quark masses are uncorrelated, so
that the error in the chiral extrapolation is larger than for quenched
QCD, probably obscuring sea-quark effects~\cite{9806027}. As
reported by CP-PACS~\cite{ruedi_talk,kanaya_talk}, the trend in the
light hadron masses, as lattice spacing decreases, is to agree with
experiment, but the statistical errors are still large compared to the
quenched QCD results, where the discrepancy with experiment is small anyway, and
CP-PACS does not attempt a continuum extrapolation.

\subsection{Staggered quarks}
\begin{figure}[htb]
\epsfxsize=0.5\textwidth
\hspace*{-3mm}\epsffile{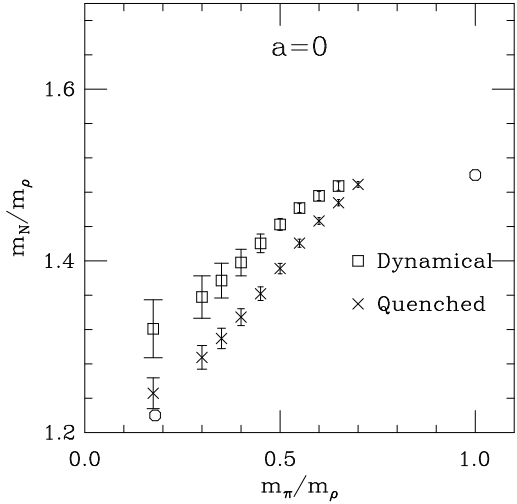}\\[-12mm]
\caption{Edinburgh plot comparing continuum results for quenched and
$N_f=2$ QCD using staggered quarks, obtained by MILC~\cite{gottlieb_poster}.}
\label{fig:milc_dyn_ed_plot}
\end{figure}
The results obtained by MILC for $N_f=2$ staggered
quarks~\cite{gottlieb_poster} are perhaps the most surprising at this
year's conference. They perform a $\mbox{const} + O(a^2)$ continuum
extrapolation of data at five lattice spacings, at fixed $m_\pi/m_\rho$,
to produce the Edinburgh plot shown in Fig~\ref{fig:milc_dyn_ed_plot}.
Although the continuum and chiral extrapolations appear to be well behaved,
the dynamical results lie systematically above the quenched results, and
so deviate further from experiment. This is true even at large quark
masses, where the entire calculation should be under control! 

\subsection{The strange-quark sector}
\begin{figure}[htb]
\epsfxsize=0.44\textwidth
\vspace*{25mm}
\hspace*{-3mm}\epsfbox[10 60 510 560]{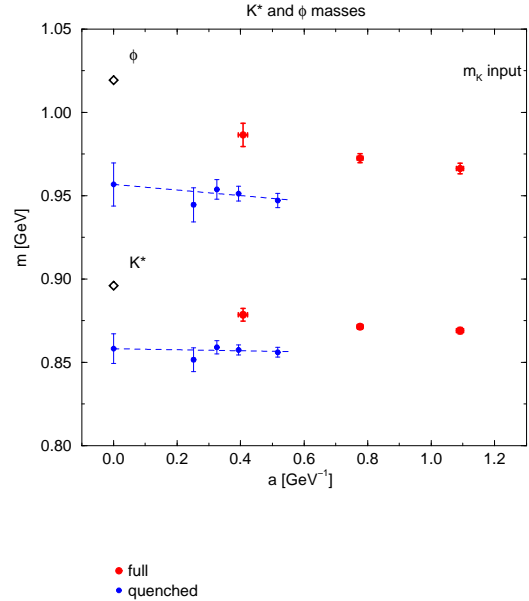}\\[-25mm]
\caption{Comparison of the strange meson splitting in quenched and
$N_f=2$ QCD, obtained by CP-PACS~\cite{ruedi_talk,kanaya_talk}.}
\label{fig:cppacs_hyperfine}
\end{figure}
Returning to the Wilson data, we might hope to see an improvement in the
strange sector from dynamical quarks. CP-PACS reports a tendency for the
strange meson splitting to increase as the continuum limit is
approached~\cite{ruedi_talk,kanaya_talk}, in better agreement with
experiment than the quenched data, as shown in
Fig~\ref{fig:cppacs_hyperfine}. 

SESAM points out that $J$ no longer has a simple interpretation as the
slope of the vector meson mass with respect to the square of the
pseudoscalar meson mass, $m_{\rm PS}$, because $J$ is no longer a
function only of $m_{\rm PS}$, but depends separately on the valence and
sea-quark masses~\cite{9806027}. Thus, chiral extrapolation cannot be
avoided in computing $J$. At fixed sea-quark mass and fixed lattice
spacing, there is no improvement in the value of $J$ compared to the
quenched approximation, and there is no discernable dependence on the
sea-quark mass, as noted also by UKQCD~\cite{9806027,talevi_talk}. This
could be due to the continued use of the quenched approximation for the
strange quark.

\subsection{Quark masses}

Partial quenching, in which the valence and sea-quark masses are taken
to be different, and which is necessary for the strange quark in $N_f=2$
simulations, allows for an alternative definition of quark mass at
non-zero lattice spacing. The partially-quenched quark mass is defined
as
\begin{equation}
m^{\rm PQ}a 
\equiv \left(\frac{1}{2\kappa_{\rm valence}}
      -\frac{1}{2\kappa_{\rm crit}(\kappa_{\rm sea})}\right)
\end{equation}
where $\kappa_{\rm crit}(\kappa_{\rm sea})$ is defined by the vanishing
of the pseudoscalar meson mass, at fixed $\kappa_{\rm sea}$. Last year,
SESAM reported that $\kappa_{\rm crit}(\kappa_{\rm sea})$ depends
strongly on $\kappa_{\rm sea}$~\cite{sesam_letter}. 

\begin{figure}[tb]
\epsfxsize=0.42\textwidth
\vspace*{22mm}
\epsfbox[10 60 510 560]{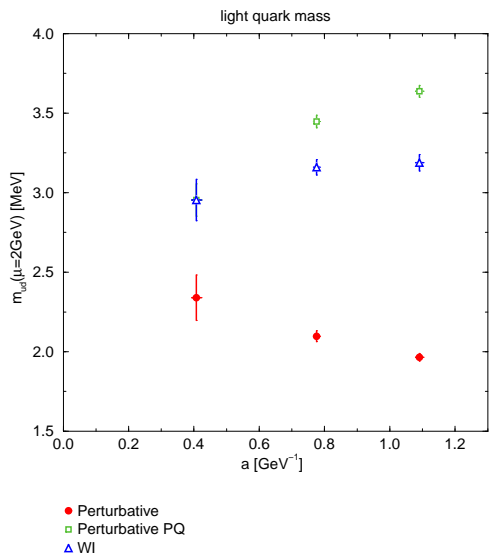}
\vspace*{-22mm}
\caption{CP-PACS results for the $ud$ quark mass in two-flavour QCD at
three lattice spacings, using the vector WI (Perturbative in the figure), axial WI
(WI) and partially-quenched (Perturbative PQ)
definitions~\cite{ruedi_talk,kanaya_talk}.}
\label{fig:cppacs_dyn_ud_mass}
\end{figure}
CP-PACS takes $\kappa_{\rm sea}=\kappa_{ud}$ to define a
partially-quenched mass.  Their results~\cite{ruedi_talk,kanaya_talk} in
Fig~\ref{fig:cppacs_dyn_ud_mass} show that, at fixed lattice spacing,
the partially-quenched, vector and axial Ward identity masses all
disagree. However, it appears that, within the relatively large
statistical errors, all definitions are converging to a continuum limit
of around 2.5~MeV, although the data is not good enough to justify an
extrapolation.
\begin{figure}[htb]
\epsfxsize=0.42\textwidth
\vspace*{22mm}
\epsfbox[10 60 510 560]{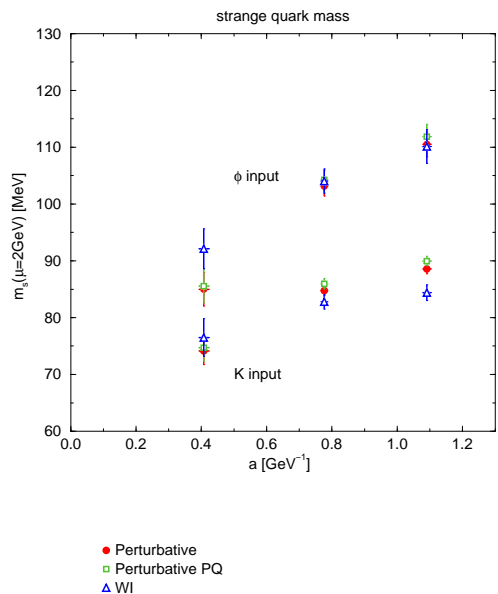}
\vspace*{-19mm}
\caption{Comparison of CP-PACS results for the $s$ quark mass,
determined from the $\phi$ and from the $K$ meson masses, in two-flavour
QCD~\cite{ruedi_talk,kanaya_talk}.}
\label{fig:cppacs_dyn_s_mass}
\end{figure}
As shown in Fig~\ref{fig:cppacs_dyn_s_mass}, CP-PACS also finds that the
estimates for the strange quark mass, obtained using the $K$ and $\phi$
masses, disagree at non-zero lattice spacing, but, within the large
statistical errors, appear to be consistent with a single continuum
limit around 80~MeV. Evidently, there is a large quenching effect which
overestimates the light quark masses by as much as 40\%. It should be
noted that the results for the strange quark mass may still be an
overestimate due to the use of partial quenching. In fact, CP-PACS
estimates for the ratio $m_s/m_{ud}$ from the axial Ward identity are
essentially independent of lattice spacing at a value of about
25~\cite{ruedi_talk,kanaya_talk}, consistent with lowest-order chiral
perturbation theory, although a slightly smaller value is favoured at
next order~\cite{leutwyler}.

\subsection{String breaking}
\begin{figure}[htb] 
\epsfxsize=0.48\textwidth 
\vspace*{11mm} 
\hspace*{-10mm}\epsfbox[10 60 510 560]{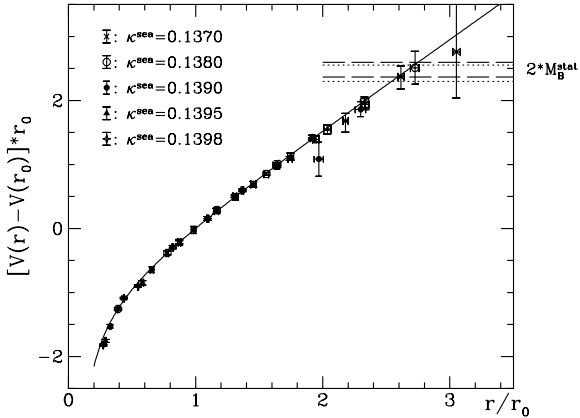}\\[-40mm] 
\caption{UKQCD results for the static quark potential in two-flavour QCD
for different sea-quark masses~\cite{talevi_talk}.} 
\label{fig:ukqcd_potential}
\end{figure}
The much-hoped-for signal of sea-quark effects is, of course,
string-breaking. Although Philipsen and Wittig have demonstrated very
clearly the crossover from a string-like to a two-meson state in a
$(2+1)$-dimensional SU(2) Higgs model~\cite{philipsen}, zero-temperature
QCD results, such as those from UKQCD in Fig~\ref{fig:ukqcd_potential},
show little dependence of the potential at large distances on sea-quark
mass~\cite{talevi_talk}. It is likely that the Wilson loop operator does
not project well onto broken string states outside a narrow mixing
region~\cite{philipsen}, and a computation of the full matrix correlator
of Wilson loops and two-meson operators is needed, but
see~\cite{kuti_talk} for further discussion of this. 

\section{IMPROVED ACTIONS}

Improvement at least to $O(a)$ is necessary for dynamical quark
simulations and considerable progress has been made in recent years,
particularly by the Alpha Collaboration, in implementing this
non-perturbatively for Wilson quarks. Most of the scaling tests have
been performed in quenched QCD, but some dynamical results are now
becoming available. Higher-order improvement has the potential for big
pay-offs, but many parameters have to be fixed reliably.

\subsection{$O(a)$-improved Wilson quarks}

This requires one additional term in the action,
\begin{equation}
c_{\rm SW}\frac{a^5}{4}\sum_x \bar{q}(x)i\sigma_{\mu\nu}F_{\mu\nu}(x)q(x),
\label{eq:clover}
\end{equation}
as noted first by Sheikholeslami and Wohlert (SW), plus explicit
determination of the mass dependence (coefficients $b_A$, \ldots) and
mixing  (coefficients $c_A$, \ldots) at $O(a)$ of composite fields:
\begin{equation}
{\cal O}^{\rm R}
= Z_{\cal O}(1+b_{\cal O}\,am_q)({\cal O} + \sum_n c_n\, a{\cal O}_n).
\label{eq:improved_op}
\end{equation}
Using Schr\"odinger functional methods at $m_q=0$, Alpha has determined
many of the coefficients for quenched QCD for $\beta\ge 6.0$ ($a\le
0.1$~fm), by imposing chiral symmetry, which is broken at $O(a)$ by the
Wilson action (for a review see~\cite{luescher}). De Divitiis and
Petronzio use the quark-mass dependence of the PCAC
relation to extract $b_A-b_P$, $b_V-b_S$, $b_m$, $Z_mZ_P/Z_A$ and
$Z_mZ_S/Z_V$~\cite{9710071}. 

The difficulty in extending this to larger lattice
spacings is the occurrence of exceptional configurations. These can be
regulated by taking $m_q\neq 0$, and the SCRI group finds that the mass
dependence of $c_{\rm SW}$ is very weak, so that for $\beta\ge
5.7$~\cite{9711052},
\begin{equation}
c_{\rm SW}
=\frac{1-0.6084\,g^2-0.2015\,g^4+0.03075g^6}{1-0.8743g^2}
\end{equation}
within 1\% of the Alpha curve for $\beta\ge 6.0$. At fixed quark mass,
defined by $m_{\rm V}/m_{\rm PS}=0.7$, at $\beta=5.7$, scaling
violations are reduced from 41\% to 3\% in $m_{\rm V}$, and from 33\% to
2\% in $m_{\rm N}$~\cite{9711052}. 

\begin{figure}[htb] 
\epsfxsize=0.53\textwidth
\vspace*{6mm}
\hspace*{-15mm}\epsfbox[10 60 510 560]{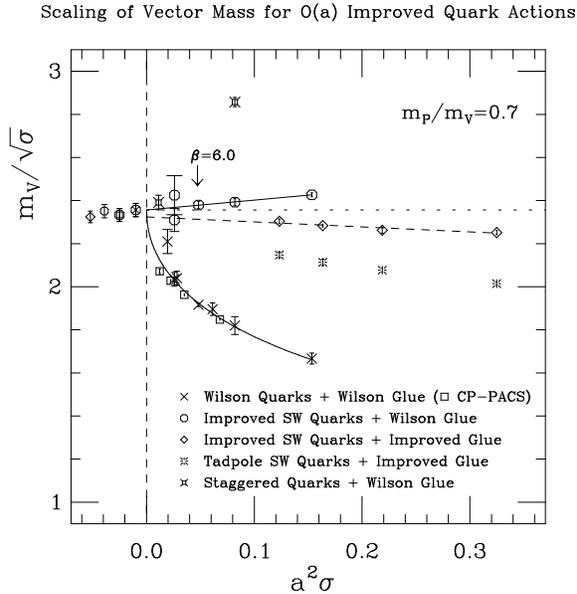}\\[-20mm] 
\caption{Comparison of scaling violations in the vector meson mass,
computed in quenched QCD using the standard Wilson action and several different $O(a)$-improved actions~\cite{klassen_talk}.}
\label{fig:scri_imp_actions}
\end{figure}

A comparison of the scaling violations in the vector meson mass, at the
same fixed quark mass, for various $O(a)$-improved actions is given in
Fig~\ref{fig:scri_imp_actions}. Consistent continuum extrapolations,
linear in $a^2$, are possible for the non-perturbatively improved
results, in agreement with $O(a)$ extrapolation of the Wilson data, but,
if the tadpole-improved results are also to be consistent, their
continuum extrapolation must include a term linear in
$a$~\cite{klassen_talk}. Also, it is interesting that the staggered
results show big $O(a^2)$ scaling violations. The QCDSF Collaboration
now has results for the non-perturbatively $O(a)$-improved Wilson
action at a smaller lattice spacing ($\beta=6.4$) than in
Fig~\ref{fig:scri_imp_actions}, which confirm the smooth approach to the
continuum limit~\cite{pleiter_talk}.

The Alpha Collaboration has computed the clover coefficient for
non-perturbative $O(a)$-improvement of two-flavour QCD with Wilson
quarks, obtaining
\begin{eqnarray}
c_{\rm SW}& = &\nonumber\\
&&\hspace*{-15mm}\frac{1-0.454\,g^2-0.175\,g^4+0.012\,g^6+0.045\,g^8}{1-0.720\,g^2}
\end{eqnarray}
valid for $\beta\ge 5.4$ and probably for $\beta\ge 5.2$~\cite{9803017}.

\subsection{Improved staggered quark actions}

The motivation for improving staggered quark actions is rather different
from that for Wilson quarks. The main objective is to reduce flavour
symmetry breaking by reducing the coupling of quarks to high-momentum
gluons. These can couple quarks at opposite corners of the Brillouin
zone, which correspond to different flavours in the continuum.
Unfortunately, there are 15 dimension-6 terms which can contribute to
the action at $O(a^2)$, making a non-perturbative estimate of their
coefficients almost impossible~\cite{luo}. One-loop perturbation theory is
typically unreliable for staggered quarks, and so the only practical
option is to use tadpole estimates. 

MILC has investigated the Naik quark action, in which a three-link
hopping term is used to cancel $O(a^2)$ terms in the staggered-quark
action at tree level, and then the coefficients are
tadpole-improved~\cite{9712010}. The gluon action they use is $O(a^2)$ one-loop
tadpole improved. Results are compared with those from staggered quarks
using both the standard, and the improved gluon action. The observed
scaling violations are $O(a^2)$ and are minimised primarily by the use
of improved glue at low quark mass, and by the use of the Naik action at
large quark mass. The speed of light from the continuum dispersion
relation is closer to unity for the Naik action than for the staggered
action (both with improved glue), whereas flavour breaking, measured
from the splitting between the non-Goldstone and the Goldstone pions, is
most sensitive to the use of improved glue, in line with the above
intuition and previous results using `fat' (ie APE smeared)
links~\cite{fat_links}. 

Laga\"e and Sinclair have generalised the MILC fat-link action,
achieving a similar reduction in flavour violation~\cite{9806014}. For
dynamical staggered quarks, MILC also finds that improving the gauge
action or fattening the links improves the flavour
symmetry~\cite{9805009}. 

Fat links can be used with the SW action to allow $c_{\rm SW}$ to be
tuned to minimise the spread of near-zero modes~\cite{9807002}. The
smeared links allow a small enough value of $c_{\rm SW}$ to avoid
exceptional configurations and, being less sensitive to UV modes of the
gauge fields, renormalisation constants are closer to 1 than for the
standard SW action.

\subsection{Higher-order improvement}

The most adventurous improved action hunters seek small scaling
violations on very coarse lattices. Inspired by the fixed-point action,
DeGrand has been testing fermion actions which couple all the fields on
a $3^4$ hypercube~\cite{9802012}. He finds rotational invariance (ie the
speed of light from the continuum dispersion relation is close to 1)
even at $a=0.36$~fm, although more complicated Pauli interactions seem
no better than the simple clover term, Eq~(\ref{eq:clover}), for boosting the hyperfine splitting. 

The $O(a^3)$-accurate D234c quark action of Alford et
al.~\cite{9712005}, with plaquette plus $2\times 1$ rectangle gluon
action and mean-link (Landau gauge) tadpole-improved coefficients, has
scaling violations of only 7\% (in $m_\phi$ at $m_{\rm PS}/m_{\rm
V}=0.7$), and a much better dispersion relation than the SW action, at a
lattice spacing as large as 0.4~fm. 

Thus, while non-perturbative $O(a)$ improvement for Wilson quarks is
well-established for quenched QCD, and so sets almost a mandatory
minimum improvement level for dynamical simulations, there are further
encouraging signs that the big gains from higher-order improved actions
could become a practical reality.

\section{NON-PERTURBATIVE QUARK\\MASS RENORMALISATION}

Most of the existing results for quark masses have used perturbative
matching and continuum extrapolation. There are some indications that
this procedure may not be fully under control. Inconsistencies between
the results from using different definitions and different actions cast
doubt on the use of perturbation theory, but the issue may be clouded by
discretisation errors. Consequently, the main progress this year has not
been to increase the precision of the mass estimates compared to last
year's review~\cite{gupta97}, but rather in implementing two
non-perturbative matching schemes. The work described below is entirely
in the quenched approximation.

\subsection{Perturbative matching}

The CP-PACS results, given in Eqs~(\ref{eq:cppacs_ms}) and
(\ref{eq:cppacs_mud}), obtained with the Wilson action and 1-loop $Z$
factors, may be compared with the QCDSF group's new results using the
non-perturbative $O(a)$-improved Wilson action~\cite{pleiter_talk}. 

QCDSF has new data at $\beta=6.2$ (on $24^3\times 48$ and $32^3\times
64$ lattices) and at $\beta=6.4$ (on a $32^3 \times 48$ lattice). They
determine the quark masses using the axial Ward identity, with
non-perturbative values of $Z_A$ from the Alpha Collaboration, and
tadpole-improved 1-loop perturbative values for $Z_P$ in
Eq~(\ref{eq:axialWI}). They fix the strange quark mass from $M_K$ and
extrapolate results at $\beta=6.0$, 6.2 and 6.4 linearly in $a^2$ to the
continuum, obtaining
\begin{eqnarray}
m_{ud}^{\overline{\rm MS}}(2\;{\rm GeV})
&=& 4.13\pm 0.08\;{\rm MeV}\\
m_s^{\overline{\rm MS}}(2\;{\rm GeV})
&=& 98.1\pm 2.4\;{\rm MeV}
\label{eq:qcdsf_ms}
\end{eqnarray}
which are lower than the CP-PACS values, and significantly so for the
strange quark.

Working at fixed lattice spacing, with tree-level $O(a)$-improved Wilson
quarks at the same $\beta$ values as QCDSF, Gim\'enez et
al.~\cite{9801028,giusti_talk} also find that the strange quark mass
defined by the axial Ward identity with perturbative matching is
significantly lower than that from the vector Ward identity, or from the
non-perturbative RI scheme (see below). However, the difference could
partly be due to discretisation effects.

Before discussing non-perturbative matching, it is interesting to note
that the first results for quark masses using domain wall fermions were
reported this year~\cite{wingate_talk}. The preliminary results, using
$M_K$ to fix the strange quark mass, $f_\pi$ to set the scale, and
perturbative matching, at $\beta=5.85$, 6.0 and 6.3, appear to scale
within relatively large errors. The weighted average is
$m_s^{\overline{\rm MS}}(2\;{\rm GeV}) = 82(15)$~MeV. While the error is
too large to add much new information at the moment, the result shows
that domain wall fermions can provide a valuable independent
determination of quark masses.

\subsection{The non-perturbative RI scheme}

In the RI scheme~\cite{RI_scheme,9801028}, renormalisation conditions
are imposed on amputated Green functions between quark states of
momentum $p$ in Landau gauge, setting them equal to their tree-level
values and thereby maintaining Ward identities. Eg for $O_\Gamma =
\overline{q}\Gamma q$, the amputated Green function is
\begin{equation}
\Lambda_O(pa) = S_q(pa)^{-1}G_O(pa)S_q(pa)^{-1}
\end{equation}
where the Green function, $G_O(pa)$, and quark propagator, $S_q(pa)$,
are computed by Monte Carlo. Then the renormalisation condition is
\begin{equation}
\left.Z_O^{\rm RI}(\mu a)Z_q^{-1}(\mu a) 
{\rm Tr}\,P_O\Lambda_O(pa)\right|_{p^2=\mu^2} = 1,
\end{equation}
where $P_O$ projects onto the tree-level amputated Green function and
$Z_q$ is the wavefunction renormalisation constant.  In this way,
gauge-dependent non-perturbative renormalisation is achieved at the
lattice scale $\mu\simeq a^{-1}\simeq 2-4$~GeV.

In order to obtain results in the  $\overline{\rm MS}$ scheme, a
continuum perturbative matching calculation is required. This has now
been done at next-to next-to-leading order (NNLO)~\cite{9803491}, ie to
$O(\alpha_s^2)$, and, when combined with the non-perturbative RI result,
cancels the gauge dependence.

As mentioned above, this method has been applied to tree-level
$O(a)$-improved Wilson quarks at $\beta=6.0$, 6.2 and 6.4, using both
the vector and axial Ward identity definitions of quark
mass~\cite{9801028,giusti_talk}. These definitions give consistent results in
the non-perturbative RI scheme, and show no lattice-spacing dependence
over the range studied (although they are not consistent with the
results of perturbative matching), so the best estimates, obtained by
averaging the data, are~\cite{giusti_talk}
\begin{eqnarray}
m_{ud}^{\overline{\rm MS}}(2\;{\rm GeV})
&=& 5.7\pm 0.5\pm 0.8\pm 0.8\;{\rm MeV}\\
m_s^{\overline{\rm MS}}(2\;{\rm GeV})
&=& 130\pm 8\pm 15\pm15\;{\rm MeV}
\end{eqnarray}
where the errors are due to statistics, non-perturbative
renormalisation and systematics, respectively. Clearly, the errors in
these estimates are too large at present to resolve the discrepancy
between Eq~(\ref{eq:qcdsf_ms}) (QCDSF) and Eq~(\ref{eq:cppacs_ms})
(CP-PACS). 

Kilcup and Pekurovsky~\cite{pekurovsky_talk} have reported preliminary
results for the strange quark mass obtained using non-perturbative RI
renormalisation for staggered quarks, extrapolated to the continuum from
data at $\beta=6.0$, 6.2 and 6.4. They obtain $m_s^{\overline{\rm
MS}}(2\;{\rm GeV}) = 129\pm 23$~MeV, in good agreement with other
estimates. 

What is needed now is a higher statistics determination,
using an improved action ($O(a)$-improved Wilson, or staggered), which
allows for a reasonably confident continuum extrapolation to compare
with the CP-PACS and QCDSF perturbatively-matched results.

\subsection{Non-perturbative running of the\\quark mass}

As reported last year~\cite{luescher97}, the Alpha Collaboration has
been implementing a non-perturbative method for running the quark mass
in the Schr\"odinger Functional (SF) scheme from low to high scales. The
$O(a)$-improved PCAC mass in the SF scheme is~\cite{wittig_talk}
\begin{equation}
\overline{m}_{\rm SF}(2L)
= \frac{Z_A(1+b_A\,am_q)}{Z_P(2L)(1+b_P\,am_q)}\;m^{\rm PCAC}_{\rm lat}
\end{equation}
where $m^{\rm PCAC}_{\rm lat}$ is defined by the ratio of matrix
elements in Eq~(\ref{eq:axialWI}), in which the axial vector current and
pseudoscalar density are improved to order $a$, as in
Eq~(\ref{eq:improved_op}). Here the scale dependence comes through
$Z_P(2L)$, which is defined in the SF scheme in a box of linear size
$2L$.

\begin{figure}[htb] 
\vspace*{3mm}
\epsfxsize=0.58\textwidth
\hspace*{8mm}\epsffile{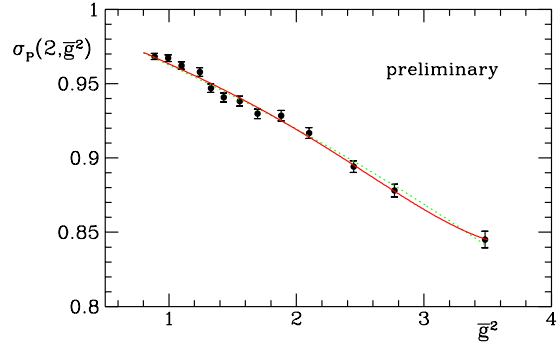}\\[-10mm] 
\caption{Step scaling function for the quark mass in quenched QCD~\cite{wittig_talk}.}
\label{fig:step_scaling_fn}
\end{figure}
Non-perturbative running is implemented via step scaling functions,
\begin{eqnarray}
\overline{g}_{\rm SF}^2(2L) 
&=& \sigma(2,\overline{g}_{\rm SF}^2(L))\\  
Z_P(2L) 
&= & \sigma_P(2,\overline{g}_{\rm SF}^2(L))Z_P(L),
\end{eqnarray}
which can be iterated to generate a sequence of couplings and quark masses,
\begin{equation}
\begin{array}{ll}
u_k=\overline{g}_{\rm SF}^2(2^kL),\hspace{5mm} & m_k=\overline{m}_{\rm
SF}(2^kL
) \\[2mm]
u_{k+1} = \sigma(2,u_k), & m_{k+1} = m_k/\sigma_P(2,u_k).
\end{array}
\end{equation}
The step scaling functions are computed for different lattice spacings
and extrapolated to the continuum limit. The step scaling function for
the quark mass is shown in Fig~\ref{fig:step_scaling_fn}.

\begin{figure}[htb] 
\vspace*{3mm}
\epsfxsize=0.58\textwidth
\hspace*{8mm}\epsffile{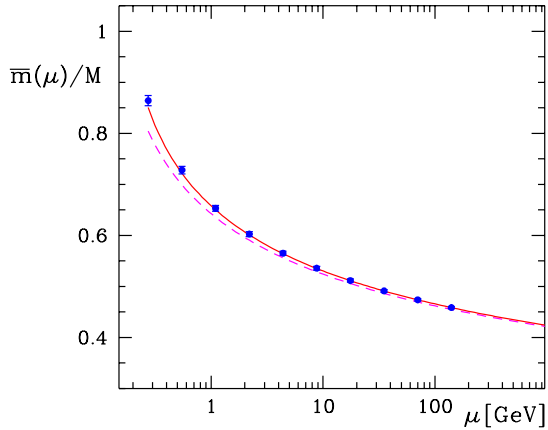}\\[-5mm] 
\caption{Universal running of the SF quark mass in quenched 
QCD~\cite{wittig_talk}.}
\label{fig:running_mass}
\end{figure}

Successive application of $\sigma_P$ yields ratios
\begin{eqnarray}
{{\overline{m}_{\rm SF}(L_{\sf max})}\over{\overline{m}_{\rm SF}(2L_{\sf
max})}}
\hspace*{-3mm}&,&\hspace*{-2mm}
{{\overline{m}_{\rm SF}(L_{\sf max}/2)}\over{\overline{m}_{\rm
SF}(2L_{\sf max})
}},\ldots\nonumber\\
&&\hspace*{8mm}\ldots,
{{\overline{m}_{\rm SF}(L_{\sf max}/2^8)}\over{\overline{m}_{\rm
SF}(2L_{\sf max
})}}.
\end{eqnarray}
At the smallest coupling, the SF quark mass can be related to the RG invariant
mass, Eq~(\ref{eq:rg_invt_mass}), through
\begin{equation}
{\overline{m}\over M} = \left(2b_0\overline{g}^2\right)^{d_0/{2b_0}}
\exp\int_0^{\overline{g}}{\rm d}g\left[{\tau(g)\over\beta(g)}
-{d_0\over{b_0g}}\right]
\end{equation}
and, hence, the evolution of the running quark mass between this
high-energy scale and the original low-energy scale, $2L_{\sf max}$, is
determined. This universal result is shown in Fig~\ref{fig:running_mass}.

\begin{figure}[htb] 
\vspace*{-26mm}
\epsfxsize=0.64\textwidth
\hspace*{8mm}\epsffile{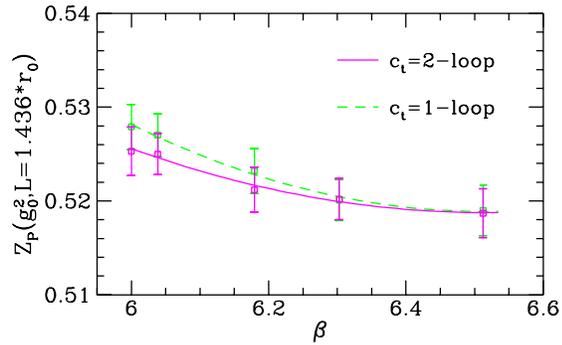}\\[5mm] 
\caption{Renormalisation factor for the pseudoscalar density in the SF
scheme at the low-energy scale $2L_{\sf max}$, as a function of the gauge
coupling in quenched QCD~\cite{wittig_talk}.}
\label{fig:Z_P}
\end{figure}

At the lowest energy, Alpha obtains
\begin{eqnarray}
2L_{\sf max} &=& 1.436\, r_0 \simeq 0.72\,{\rm fm}\\
{M\over\overline{m}_{\rm SF}(2L_{\sf max})} &=& 1.166\pm 0.015\pm 0.008.
\end{eqnarray}
So the total renormalisation factor is
\begin{equation}
{M\over{m_{\rm lat}}} = 1.166(17)\,
{{Z_A(g_0)(1+b_Aam_q)}\over{Z_P(g_0,2L_{\sf max})(1+b_Pam_q)}}
\end{equation}
where $b_A-b_P$ is small enough to be safely neglected~\cite{9710071}.
The matching to the lattice scheme is completed by computing
$Z_P(g_0,2L_{\sf max}=1.436r_0)$ for a range of $\beta$ values, by
finding $(\beta,L/a)$ so that $L/a=1.436\,r_0/a$, using known
values of $r_0/a$~\cite{9806005}. The results are shown in
Fig~\ref{fig:Z_P}, where the two sets of results correspond to using the
1- and 2-loop estimates for the boundary counterterm, $c_t$, which is
only known perturbatively. 
All that is required now, for example, is the PCAC mass, $(m_u+m_s)_{\rm
lat}$, from a standard simulation, to obtain the RG invariant mass
$M_u+M_s$. Currently, this last step is missing, but suitable data have
been generated by both QCDSF and UKQCD.

\section{GLUEBALLS \& HYBRIDS}

Finally, I will turn to phenomenology. Quenched QCD predictions are
playing an important part in glueball and hybrid meson
searches~\cite{close97}. The challenge is to quantify mixing and
sea-quark effects.

\subsection{The lightest glueball}
\begin{figure}[htb] 
\vspace*{-3mm}
\epsfxsize=0.5\textwidth
\hspace*{-3mm}\epsffile{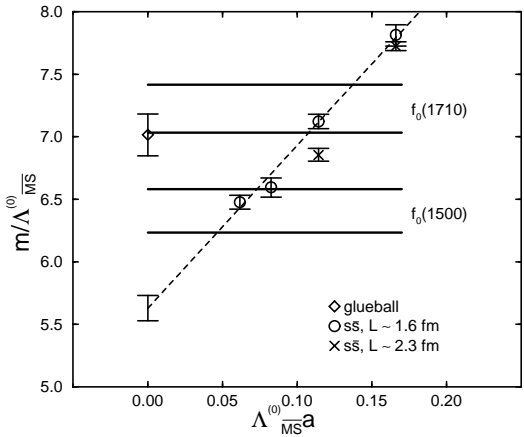}\\[-15mm] 
\caption{Continuum extrapolation of the unmixed scalar $s\bar{s}$ mass,
compared with the continuum limit of the scalar glueball in quenched QCD
and the masses of the $f_0(1710)$ and the $f_0(1500)$ (the error bands
are due to the uncertainty in $\Lambda^{(0)}_{\overline{\rm
MS}}$)~\cite{9805029}.} 
\label{fig:glueballs}
\end{figure}
Lee and Weingarten~\cite{9805029} have estimated the mixing of the
lightest discrete isosinglet states. Their results for the unmixed
scalar quarkonium mass are shown in Fig~\ref{fig:glueballs}, together
with their best estimate of the $O^{++}$ glueball mass, from world data.
From the relative ordering of the continuum glueball and $s\bar{s}$ meson, it
is plausible that the $f_0(1710)$ is predominantly glue. Lee and
Weingarten have computed the mixing energy, and use it to fit the
experimental data to a crude model of the mixing between the glueball,
and the scalar $s\bar{s}$ and $(u\bar{u} + d\bar{d})/\sqrt{2}$ (which
they denote $n\bar{n}$) states, with the result
\begin{eqnarray}
|f_0(1710)\rangle & = & 0.86(5)|g\rangle
                        + 0.30(5)|s\overline{s}\rangle\nonumber\\
                  &   & + 0.41(9)|n\overline{n}\rangle\\  
|f_0(1500)\rangle & = & - 0.13(5)|g\rangle
                        + 0.91(4)|s\overline{s}\rangle\nonumber\\
                  &   & - 0.40(11)|n\overline{n}\rangle\\ 
|f_0(1390)\rangle & = & - 0.50(12)|g\rangle
                        + 0.29(9)|s\overline{s}\rangle\nonumber\\
                  &   & + 0.82(9)|n\overline{n}\rangle,
\end{eqnarray}
which points to the $f_0(1710)$ being 74\% glueball. They argue
that, despite being largely $s\bar{s}$, the opposite sign of
$|s\overline{s}\rangle$ and $|n\overline{n}\rangle$ in
$|f_0(1500)\rangle$ suppresses its decay to $K\overline{K}$, in line
with the experimental observation.

\subsection{Hybrid mesons}

The UKQCD quenched QCD result~\cite{ukqcd_exotic} that the lightest
$s\bar{s}$ exotic meson has $J^{PC}=1^{-+}$ and a mass of 2.0(2)~GeV, is
supported by further quenched results reported by MILC this
year~\cite{mcneile_poster}. SESAM has repeated the UKQCD analysis using
Wilson $N_f=2$ configurations at $\beta=5.6$. They are able to perform a
linear chiral extrapolation in the dynamical quark mass, from data at
four quark masses, and find that the $1^{-+}$ remains the lightest
exotic, with a mass of 1.9(2)~GeV, which is consistent with the quenched
results~\cite{lacock_poster}.  Of course, exotic mesons are expected to
mix with four-quark states. This mixing is currently under investigation
by SESAM.

For the wider particle physics community, predictions for glueballs and
hybrid mesons (along with the light quark masses) are the most
interesting output of our light hadron work, and pursuing a more
systematic analysis of mixing and sea-quark effects will be an important
part of our future programme.

\section{CONCLUSIONS \& ACKNOWLEDGEMENTS}

In summary, we now have precise results from CP-PACS which indicate
that quenched QCD fails to reproduce the light hadron spectrum at the
few percent level. Even so, there may be lingering doubts about the
reliability of the continuum extrapolation, as there is no confirmation
yet from improved actions. For quark masses, we are on the verge of
resolving discrepancies between different simulations, which are
probably due to the unreliability of perturbative matching. Techniques
for non-perturbative renormalisation are well-developed and should be
mandatory from now on. It is time for a definitive simulation of quenched
QCD using non-perturbative $O(a)$-improved Wilson, or staggered quarks
and non-perturbative matching, to establish the quenched light hadron
spectrum and quark masses once and for all. 

Simulations with two flavours of dynamical quarks are showing hints that
the strange hadron spectrum is in better agreement with experiment than
in quenched QCD, although the quenched approximation must still be used
for the strange quark. The biggest effect observed so far is that the
light quark masses are around 40\% smaller than in quenched QCD. String
breaking has not been seen yet, but it should be within reach of present
day simulations, and there is much to be done to improve our
understanding of mixing and sea-quark effects on glueball and hybrid
states.

Somewhat surprisingly, quenched QCD has turned out to be a rather
accurate effective theory for many quantities. Knowing its limitations
should give us greater confidence in using it as a phenomenological
model for quantities where 10-20\% accuracy is useful. To improve on it,
will require simulations with several different combinations of
sea-quark flavours, in order to be able reliably to
interpolate/extrapolate to the flavour content of the real world. We have 
reached the point where quenched QCD can be simulated with sufficient
accuracy and this gives us a solid foundation from which to tackle the flavour dependence of QCD.

I wish to thank M.~Alford, R.~Burkhalter, L.~Giusti, M. Golterman,
S.~Gottlieb, S.~G\"usken, T.~Klassen, P.~Lacock, K.-F.~Liu, X.-Q. Luo,
T.~Mendes, C.~McNeile, T.~Okude, S.~Ohta, E.~Pallante, G.~Schierholz,
D.~Sinclair, A.~Soni, A.~Vladikas, D.~Weingarten, M.~Wingate, H.~Wittig,
and T.~Yoshi\'e for sending me their results in advance.

\end{document}